\documentclass[aps,prb,twocolumn,showpacs]{revtex4}

\usepackage{graphicx}
\newcommand{\nn}{\nonumber}
\newcommand{\al}{\alpha}
\newcommand{\be}{\beta}
\newcommand{\la}{\langle}
\newcommand{\ra}{\rangle}

\begin{document}

\title{Electron-nuclei spin relaxation through phonon-assisted hyperfine interaction \\
in a quantum dot}%

\author{Veniamin A. Abalmassov}%
\email{V.Abalmassov@isp.nsc.ru}%
\affiliation{Institute of Semiconductor Physics and Novosibirsk
State University, \\ SB RAS, 630090 Novosibirsk, Russia}
\author{Florian Marquardt}
\email{Florian.Marquardt@yale.edu}%
\affiliation{Department of Physics, Yale University, New Haven CT, 06511, USA}%


\date{\today}%

\begin{abstract}
We investigate the inelastic spin-flip rate for electrons in a
quantum dot due to their contact hyperfine interaction with
lattice nuclei. In contrast to other works, we obtain a
spin-phonon coupling term from this interaction by taking directly
into account the motion of nuclei in the vibrating lattice. In the
calculation of the transition rate the interference of first and
second orders of perturbation theory turns out to be essential. It
leads to a suppression of relaxation at long phonon wavelengths,
when the confining potential moves together with the nuclei
embedded in the lattice. At higher frequencies (or for a fixed
confining potential), the zero-temperature rate is proportional to
the frequency of the emitted phonon. We address both the
transition between Zeeman sublevels of a single electron ground
state as well as the triplet-singlet transition, and we provide
numerical estimates for realistic system parameters. The mechanism
turns out to be less efficient than electron-nuclei spin
relaxation involving piezoelectric electron-phonon coupling in a
GaAs quantum dot.
\end{abstract}


\maketitle

\section{Introduction}

Electron spin dynamics in mesoscopic devices has been attracting a
lot of attention recently in the context of spintronics
\cite{Wolf} and quantum computation \cite{NCh, DV}. A crucial
feature of this dynamics is the relaxation of the electron's spin
due to the interaction with an environment. Generally speaking,
the coherence of an electronic spin state vanishes during the time
$T_2$, which limits the possibility of coherent manipulation of
qubits, while relaxation to thermal equilibrium occurs during
another time $T_1$, which is usually larger than $T_2$
\cite{Abragam}.

Several types of spin-dependent interactions can give rise to
electron spin relaxation, e.g. the electron spin-orbit interaction
\cite{PTreview, Hasegawa, Roth, Frenkel, KhN1, KhN2, MKGB, WRL-G,
Tahan, Glavin} and the electron-nuclei hyperfine interaction
\cite{Abragam, DPreview, PBS, KimVX, ENF, EN, Lyanda-Geller,
Sousa}. Their action depends essentially on the dimension of the
system. Systems of zero dimension, i.e. quantum dots (QDs), are
characterized by a discrete electron energy spectrum. In this case
energy conservation in the spin-flip process usually can be
fulfilled only by transferring the energy to another subsystem,
e.g. phonons. The energy transfer includes both the Zeeman energy
of the electron spin in an external field and possibly the energy
of an orbital transition. A discussion of other electron spin
relaxation mechanisms not mentioned above and relevant to QDs can
be found in Ref. \onlinecite{KhN2}, for example. Electron spin
relaxation in a QD due to hyperfine interaction alone, in the absence of an external magnetic field,
was investigated recently in Refs.
\onlinecite{MER, SchKhL}.

Many years ago, the phonon-assisted electron spin-flip transition
between Zeeman sublevels due to hyperfine interaction with an
impurity nucleus and lattice nuclei was considered for
impurity-bound electrons in silicon \cite{PBS}, where the authors
investigated the nuclear polarization in the Overhauser effect.
The process that has been considered in that paper is associated
with a crystal dilation and a corresponding adiabatic change in
the electron effective mass and, as a consequence, in the electron
envelope wave function and the hyperfine coupling constant.
Recently, electron spin relaxation due to the hyperfine contact
interaction has been readdressed with an application to GaAs QDs
\cite{ENF, EN}. The transition amplitude was calculated in
second-order perturbation theory, describing the action of the
hyperfine contact interaction and a spin-independent piezoelectric
electron-phonon coupling.

In this paper we will analyze another spin relaxation mechanism
provided by the combination of hyperfine contact interaction and
the influence of phonons on the electron inside a QD. In our
approach we take into account directly the phonon-induced motion
of nuclei which are coupled to the electron spin through the
hyperfine interaction. The electron-phonon interaction appears via
the displacement field shifting the positions of the nuclei, and
therefore is independent of the piezoelectric coupling that
applies only to crystals without inversion symmetry (such as GaAs,
but not Si). Moreover, this mechanism allows the electron-nucleus
spin flip-flop with a simultaneous emission of a phonon to appear
already in first-order perturbation theory. Nevertheless, it will
turn out that it is necessary to keep as well the second-order
terms associated with the motion of the electron confining
potential, since they lead to a crucial cancellation of
first-order terms at low frequencies of the emitted phonon,
thereby suppressing the relaxation rate. The physical reason
behind this is the following: Long-wavelength phonons displace
both the lattice nuclei and the electron's confining potential in
the same way. However, it is only the relative motion of the
electron with respect to the nuclei that enables a transition.
Therefore, the influence of long-wavelength phonons is suppressed.
If, on the other hand, the confining potential can be considered
as fixed or it moves independently from the lattice nuclei, this
suppression does not apply any more, and the destructive
interference between first-order and second-order terms is broken.

The present article is organized as follows: In Section
\ref{model}, we introduce the Hamiltonian of our model, including
the effect of lattice vibrations on the hyperfine coupling and the
confining potential. After that, we derive the electron spin flip
transition amplitudes due to this perturbation, discussing the
partial cancellation of terms and the dominating contribution. We
calculate the transition rate, Eq. (\ref{zeeman-t1}), for the case
of Zeeman-split sublevels of the electron ground state (Sec.
\ref{zeemansec}), and perform an analogous derivation for the case
of a triplet-singlet transition (Sec. \ref{tripletsec}). Finally,
in Section \ref{numestsec} we look at numerical estimates for
realistic system parameters.

\section{The model}
\label{model}

In our model, we will assume the displacement of the QD confining
potential to be described by the phonon displacement field
evaluated at the center of the dot (which we take to be the
origin). Note that this is analogous to the case of an
impurity-bound electron. Any more
detailed description (e.g. allowing for a distortion of the
potential) would require further specifications concerning the way
this potential is applied to the dot, but would not add
significantly to the realism of the present model.

In the effective mass approximation the Hamiltonian of the system
of electrons and phonons in the QD has the following form, if the
perturbations due to electron-phonon and hyperfine interactions
are excluded:
\begin{eqnarray}\label{hamiltonian-zero}
    \!\!\hat{H}_{0} \!\!&=&\! \sum_{i}\left[
    \frac{\hat{{\bf P}}_{i}^2}{2m^*} +
    V({\bf r}_{i})
    + \text{g}^* \mu_{B}{\hat{\bf S}}_{i}\cdot{\bf B}\right]
    \nn \\
    &+& \!\! \frac{1}{2}\sum_{i\neq i'} V_{e-e}({\bf r}_{i}-{\bf
    r}_{{i}'}) + \!\! \sum_{\bf{k}, \lambda}
    \hbar \omega_{\bf{k},\lambda}\!
    \left[\hat{b}^\dag_{\bf{k},\lambda} \hat{b}_{\bf{k},\lambda}
    + \frac{1}{2}\right]\!,
\end{eqnarray}
where $\hat{{\bf P}}_{i}= - \text{i}\hbar {\bf \nabla}_{{\bf
r}_{i}} + e {\bf A}({\bf r}_{i})/c$ is the kinematical momentum
operator. This Hamiltonian describes interacting electrons with
effective mass $m^*$ and effective g-factor g$^*$, localized in a
static external potential $V({\bf r})$, in the presence of a
magnetic field ${\bf B}$ (with a corresponding vector potential
${\bf A}({\bf r})$), and free phonons. We neglect the Zeeman
splitting of nuclear spin states, given the small value of the
nuclear magnetic moment.

The hyperfine contact interaction of electrons of spin ${\bf
\hat{S}}_{i}$ at positions ${\bf r}_{i}$ with nuclei of spin ${\bf
\hat{I}}_{j}$ at positions ${\bf R}_{j}$ has the form
[\onlinecite{Abragam}]:
\begin{eqnarray}\label{hyperfine-interaction}
    \hat{V}_{H\!F} ({\bf r}_{i})
    &=& \sum_{j} \hat{V}_{h\!f}({\bf r}_{i}\!-\!{\bf R}_{j}) \nn\\
    &=& A v_0 \sum_{j}\hat{{\bf S}}_{i}\cdot
    \hat{{\bf I}}_{j} \; \delta ({\bf r}_{i}-
    {\bf \hat R}_{j}),
\end{eqnarray}
where the hyperfine coupling constant is determined as
\begin{equation}\label{hf-coupling-const}
    A = \frac{4 \mu_0}{3I} \frac{\mu_B \mu_I}{v_0} \,\eta ,
\end{equation}
with $\eta =|u(0)|^2$ being the square of the Bloch amplitude at
the site of the nucleus \cite{DPreview},  $\mu_B>0$ the Bohr
magneton, $\mu_I$ the nuclear magnetic moment, and $v_0$ the size
of the unit cell. The factor $\eta$, which is usually of the order
of $10^2$ to $10^3$, depending on the material, makes the Fermi
contact term much more efficient than the other terms of the
electron-nucleus hyperfine interaction \cite{Abragam}. For a more
complicated lattice structure the hyperfine constant $A$ is
defined as a sum of the individual constants $A_{j}$ over all the
nuclei in the unit cell, taking into account different values of
spins and parameters $\eta$ for different nuclei \cite{DPreview}.

The positions of the nuclei deviate slightly from equilibrium, due
to the lattice vibrations: ${\bf \hat R}_{j} = {\bf R}^{0}_{j} +
{\bf \hat u}({\bf R}^{0}_{j})$. The lattice displacement field
${\bf \hat u}({\bf R}^{0}_{j})$ is described via the phonon
creation-annihilation operators \cite{LL5},
\begin{equation}\label{displacement}
    \hat{{\bf u}}({\bf R}^{0}_{j})=\sum_{\bf{k}, \lambda}
    \sqrt{\frac{\hbar}{2\rho \, \omega_{\bf{k},\lambda} V_{ph}}}
    (\hat{b}_{\bf{k},\lambda}+ \hat{b}^\dag_{\bf{-k},\lambda}) \;
    {\bf \epsilon}_{\bf{k},\lambda} \; e^{\text{i}{\bf k}\cdot{\bf
    R}^{0}_{j}},
\end{equation}
where ${\bf \epsilon}_{\bf{k},\lambda}$ is the polarization vector
of a phonon with wave vector ${\bf k}$ in branch $\lambda$,
$\omega_{\bf{k},\lambda}$ is its frequency, $V_{ph}$ is the volume
of the crystal in which phonon modes are quantized, and $\rho$ is
the crystal mass density.

As a consequence, the total Hamiltonian acquires the following
term which can lead to a nucleus-electron spin flip-flop process
combined with the emission of a phonon:
\begin{equation}\label{perturbation-hf-phon}
    \hat{V}_{H\!F-ph}({\bf r}_{i})
    = - A v_0 \!\sum _{j}\hat{{\bf S}}_{i}\cdot
    \hat{{\bf I}}_{j}  \left(\hat{{\bf u}}({\bf R}^{0}_{j})
    \cdot \nabla_{{\bf r}_{i}}\right) \delta ({\bf r}_{i}-
    {\bf R}^{0}_{j}).
\end{equation}
In the following, we will omit the index $0$ for nuclear
equilibrium positions.

Likewise, the vibrations of the confining potential are described
by
\begin{eqnarray}\label{perturbation-deltaV}
    &&\!\!\!\! \sum_{i} \delta  \hat{V}({\bf r}_{i})
    = - \!\sum_{i} \left(\hat{{\bf u}}(0)\cdot
    {\bf \nabla}_{{\bf r}_{i}}\right) V({\bf r}_{i}) \nn \\
    \!\!\!&=& \!\! \hat{{\bf u}}(0)\,[H_{0},\sum_{i}
    \frac{\text{i}}{\hbar}\hat{{\bf P}}_{i}]
    = - \frac{m^{*}}{\hbar^2} \hat{{\bf u}}(0)
    \!\sum_{i} \left[ H_0, [H_0, {\bf r}_{i}]\right].
\end{eqnarray}
We note that the total electron momentum commutes with the
electron-electron interaction potential.

Thus, the total perturbation to the Hamiltonian $\hat{H}_0$, Eq.
(\ref{hamiltonian-zero}), is given by the three
terms described above:
\begin{equation}\label{perturbation}
    \delta \hat{H} = \sum_{i}\left[\delta \hat{V}({\bf r}_{i})
    + \hat{V}_{H\!F}({\bf r}_{i}) + \hat{V}_{H\!F-ph}({\bf
    r}_{i})\right].
\end{equation}

\section{Transitions between Zeeman sublevels}
\label{zeemansec}

At first, we will consider the transition of a single electron
between the Zeeman-split spin levels of the QD ground state.

The initial state of the system is given by the direct product of
electron, nuclear and phonon states:
\begin{eqnarray}\label{zeeman-initial}
    \vert i\ra = \vert i_e\ra\otimes\vert i_N\ra\otimes\vert i_{ph}\ra.
\end{eqnarray}
Let ${\bf n}$ denote the direction of magnetic field. The initial
electron state $\vert i_e\ra = \vert \psi({\bf r})\ra\otimes \vert
{\bf n}_{-}\ra$ is given by the product of the spin wave function
$\vert {\bf n}_{-}\ra$, which is an eigenfunction of the equation
$({\bf n}\cdot \hat{{\bf S}})\vert {\bf n}_{\pm}\ra = \pm
\frac{1}{2}\vert {\bf n}_{\pm}\ra$, and the properly normalized
ground state coordinate wave function
\begin{eqnarray}\label{zeeman-initial-coordinate}
    \psi({\bf r}) = \frac{1}{\sqrt{V}}
    \varphi_0({\bf r}),\;\;\;\;\;\;\;
    \frac{1}{V} \int  d^3{\bf r} |\varphi_0 ({\bf r})|^2 =1.
\end{eqnarray}
Here $V$ is the effective volume of the dot. In GaAs, where the
electron g-factor is negative, the state $\vert {\bf n}_{-}\ra$
corresponds to the maximum of energy. The initial nuclear spin
state $\vert i_N\ra$ is a direct product of states of all
individual nuclei. We will average over the initial phonon field state
in the end, by inserting mean phonon occupation numbers given
by the Bose distribution function $n_\omega =1/(e^{\hbar\omega /k_B T}-1)$.

In the final state
\begin{equation}\label{zeeman-final}
    \vert f\ra = \vert f_e\ra\otimes\vert f_N\ra\otimes\vert f_{ph}\ra,
\end{equation}
the electronic spin points into the opposite direction, $\vert
f_e\ra = \vert\psi({\bf r})\ra\otimes \vert {\bf n}_{+}\ra$.
Nuclear, $\vert f_N\ra$, and phonon, $\vert f_{ph}\ra$, final
states are determined by the action of the perturbation potential
$\hat{V}_{H\!F-ph}$ which changes the electron and nuclear spin
states while conserving the total spin of the electron-nuclei
system (flip-flop process) and creates a phonon with energy
corresponding to the energy difference between electron initial
and final states.

The corresponding transition matrix element is provided by
first order perturbation theory in the potential
$\hat{V}_{H\!F-ph}({\bf r})$:
\begin{eqnarray}\label{zeeman-matrix-el-V1}
    \!\!\!\!\!\!\la \!\!\! &f& \!\!\!\! \vert \delta{\hat H} \vert
    i\ra^{(1)} \nn\\
    \!\!\!\!\!\!&=& \!\!\! - \!\sum_{j}\la f_{ph}\vert
    \hat{\bf u}({\bf R}_{j}) \vert i_{ph}\ra
    \la f_{e,N}\vert \nabla _{\bf r}
    \hat{V}_{h\!f}({\bf r}\!-\!{\bf R}_{j}) \vert i_{e,N}\ra .
\end{eqnarray}

It is important that an alternative process is possible for the
transition between the same two states, where the remaining two
terms in $\delta{\hat H}$ (namely $\delta{\hat V}$ and ${\hat
V}_{HF}$) contribute in second order perturbation theory, yielding
an amplitude that is  of the same order of magnitude as Eq.
(\ref{zeeman-matrix-el-V1}):
\begin{eqnarray}\label{zeeman-matrix-el-V2}
    \la f \vert \delta{\hat H} \vert i\ra^{(2)} &=&{\sum_{m}}'
    \left[\frac{\la f \vert \delta \hat{V}({\bf r}) \vert m\ra
    \la m \vert \hat{V}_{H\!F}({\bf r}) \vert i\ra}
    {E_{i}^{(e)} - E_{m}^{(e)}} \right. \nn \\
    &+& \left. \frac{\la f \vert  \hat{V}_{H\!F}({\bf r})\vert m\ra
    \la m \vert \delta \hat{V}({\bf r}) \vert i\ra}
    {E_{f}^{(e)} - E_{m}^{(e)}} \right],
\end{eqnarray}
where the sum is over all intermediate states which differ from
the initial and final states, and $E^{(e)}$ refers to electron
energies only. In writing down the energy denominators we have
used the fact that the hyperfine perturbation only changes the
electronic energies, and that initial and final total energies
will be the same. Note that the complete electronic energy
includes the Zeeman energy as well, and the difference between
initial and final electron energies is accounted for by the energy
of the emitted phonon, $\hbar \omega = E_{i}^{(e)} - E_{f}^{(e)}$.

Now we employ the expressions for the perturbation $\delta
\hat{V}({\bf r})$, Eq. (\ref{perturbation-deltaV}), in order to
rewrite the previous formula in a form that displays the relation to
the first order amplitude:
\begin{eqnarray}\label{zeeman-matrix-el-V2+}
    \la \!\!\! &f& \!\!\!\! \vert \delta{\hat H} \vert i\ra^{(2)}
    = \la f_{ph}\vert \hat{\bf u}(0) \vert i_{ph}\ra
    \left\{ \la f_{e,N}\vert \nabla _{\bf r}
    \hat{V}_{H\!F}({\bf r}) \vert i_{e,N}\ra \right.
    \nn \\
    \!\!\! &-& m^{*}\omega ^2 {\sum_{m}}'\left[
    \frac{\la f_{e,N}\vert \,{\bf r}\,
    \vert m_{e,N}\ra
    \la m_{e,N}\vert \hat{V}_{H\!F}({\bf r}) \vert i_{e,N}\ra}
    {E_{i}^{(e)} - E_{m}^{(e)}} \right.
    \nn \\
    \!\!\! &+& \left.\left. \frac{\la f_{e,N}\vert
    \hat{V}_{H\!F}({\bf r})\vert m_{e,N}\ra
    \la m_{e,N}\vert \,{\bf r}\,
    \vert i_{e,N}\ra}
    {E_{f}^{(e)} - E_{m}^{(e)}}\right]\right\},
\end{eqnarray}
We have used the expressions for $\delta \hat{V}({\bf r})$ in
terms of the commutator of $\hat H_0$ with both the kinematical
momentum and the position operator, as well as the fact that $\hat
V_{HF}$ commutes with the position.

The total amplitude of the transition is the sum of the terms
(\ref{zeeman-matrix-el-V1}) and (\ref{zeeman-matrix-el-V2+}). We
will regroup it into two parts. The first one consists of
Eq. (\ref{zeeman-matrix-el-V1}) and a contribution of a similar form,
the first term of Eq. (\ref{zeeman-matrix-el-V2+}),
\begin{equation}\label{zeeman-a1}
    A_1= \!\sum_{j}\la f_{ph}\vert
    \hat{\bf u}(0)-\hat{\bf u}({\bf R}_{j}) \vert i_{ph}\ra
    \la f_{e,N}\vert \nabla _{\bf r}
    \hat{V}_{h\!f}({\bf r}\!-\!{\bf R}_{j}) \vert i_{e,N}\ra.
\end{equation}
It contains the difference between the phonon displacement fields
evaluated at the origin and at the nucleus position, respectively,
which is analogous to taking the divergence of the displacement
field. This difference vanishes as $\vert \,{\bf k}{\bf R}_j
\vert$ for long-wavelength phonons and describes the important
cancellation which is only found if first and second orders of
perturbation theory are combined properly.

The second part,
\begin{eqnarray}\label{zeeman-a2}
    A_2 &=& - \la f_{ph}\vert \hat{\bf u}(0) \vert i_{ph}\ra \nn\\
    &\times& m^{*}\omega ^2 {\sum_{m}}'\left[
    \frac{\la f_{e,N}\vert \,{\bf r}\,
    \vert m_{e,N}\ra
    \la m_{e,N}\vert \hat{V}_{H\!F}({\bf r}) \vert i_{e,N}\ra}
    {E_{i}^{(e)} - E_{m}^{(e)}} \right.
    \nn \\
    \!\!\! &+& \left.\left. \frac{\la f_{e,N}\vert
    \hat{V}_{H\!F}({\bf r})\vert m_{e,N}\ra
    \la m_{e,N}\vert \,{\bf r}\,
    \vert i_{e,N}\ra}
    {E_{f}^{(e)} - E_{m}^{(e)}}\right]\right\},
\end{eqnarray}
contains a sum over intermediate states. For a single electron
making a transition between the Zeeman sublevels of its ground
state, we can set
$E^{(e)}_i-E^{(e)}_m=\varepsilon_0-(\varepsilon_m-\hbar\omega)$ in
the first summand and
$E^{(e)}_f-E^{(e)}_m=(\varepsilon_0-\hbar\omega)-\varepsilon_m$ in
the second. Note that in the two sums in Eq. (\ref{zeeman-a2})
intermediate states with the same orbital energies $\varepsilon_m$
differ by the Zeeman energy.

In order to render the following discussion concrete, we will now
specify the confining potential explicitly. We consider a QD which
is formed in a two-dimensional electron gas (2DEG) by an external
symmetric parabolic potential. The confining potential in {\it
z}-direction is usually modelled by a square well in vertical QDs
and by a triangular-shaped potential in lateral QDs. We neglect
the contributions from higher excited states in the $z$-potential,
 given their large energetic separation, and restrict our discussion to the
ground state $\chi_0(z)$. In the presence of an external magnetic
field perpendicular to the $x-y$-plane, the electron wave
functions in the lateral dimension become the Darwin-Fock
solutions (see e.g. Ref. \onlinecite{Kouwenhoven}), with the
effective confining frequency $\omega_0 = \sqrt{\Omega_0^2 +
\omega_c^2/4}$, where $\Omega_0$ is the strength of the parabolic
potential and $\omega_c = e B_{\perp}/(m^{*}c)$ the cyclotron
frequency in an external magnetic field with the component
$B_{\perp}$ perpendicular to the layer. The energy spectrum for
these states is $\varepsilon_{n \lambda} = (2n + \vert
\lambda\vert + 1) \hbar \omega_0 - \lambda\hbar \omega_c/2$.

In a harmonic oscillator the coordinate vector induces transitions
only between nearest orbital levels. Thus, the sum in Eq.
(\ref{zeeman-a2}) reduces to two terms only corresponding to the
transition from the ground state $\phi_{0 0}$ to states $\phi_{n
\lambda}$, with $n=0$ and $\lambda=\pm 1$ being the radial and the
angular momentum quantum numbers, respectively. For example, in
cylindrical coordinates we rewrite the scalar product of the
phonon polarization vector and the coordinate vector as ${\bf
\epsilon}\,{\bf r} = (\epsilon_{+}\rho e^{-\text{i}\varphi} +
\epsilon_{-}\rho e^{\text{i}\varphi})/\sqrt{2}$ and note that
$\phi_{0 0}\rho e^{\pm \text{i}\varphi}/\sqrt{2} = l \,\phi_{0
\,\pm 1}$, where $l = \sqrt{\hbar / (m^{*}\omega_0)}$ is the
length scale which determines the spatial extent of the electron
wave function in the parabolic well in the presence of an external
perpendicular magnetic field.

We now compare the two parts of the total amplitude that are given
by Eqs. (\ref{zeeman-a1}) and (\ref{zeeman-a2}). First, we note
that the matrix element $\la f_{e,N}\vert \nabla _{\bf r}
\hat{V}_{h\!f}({\bf r}\!-\!{\bf R}_{j}) \vert i_{e,N}\ra$ in the
expression for $A_1$ is proportional to $\vert \nabla_{{\bf
R}_{j}} \phi_0^2 ({\bf R}_{j})\vert$, where the gradient in the
{\it z}-direction can be estimated as $\sim 1/z_0$, with $z_0$ the
transverse dimension of the dot. Gradients in the lateral
directions are smaller by a factor $z_0/l \ll 1$.  The Zeeman
splitting, $\hbar\omega=\vert \text{g}^* \mu_{B} B \vert$, is much
less than the orbital energy splitting, $\hbar\omega \ll
\hbar\omega_0$, in typical experiments on spin relaxation in QDs.
Thus, up to a common prefactor, we can use the following estimates
for the expressions of Eqs. (\ref{zeeman-a1}) and
(\ref{zeeman-a2}):
\begin{equation}\label{A1andA2}
    \vert A_1 \vert \propto \frac{\min[k l, 1]}{z_0} \sim
    \frac{\min[\omega l/s, 1]}{z_0},
    \;\;\;
    \vert A_2 \vert \propto \frac{m^{*} \omega^2 l}{\hbar \omega_0}
    \sim \frac{\omega^2}{\omega_0^2 l},
\end{equation}
where $s$ is the mean sound velocity. Finally, we find that the
ratio $\vert A_2/A_1 \vert$ is suppressed by a factor whose form
depends on the emitted phonon wave-length:
\begin{eqnarray}\label{A1toA2}
    &&\frac{z_0}{l}\frac{\omega}{\omega_0}
    \frac{s}{\omega_0 l} \ll 1 \;\;\;\;\text{for}\;\;\;\; k l \ll 1 \nn\\
    &&\frac{z_0}{l}\left(\frac{\omega}{\omega_0}\right)^2 \ll 1
    \;\;\;\;\text{for}\;\;\;\; k l \gg 1.
\end{eqnarray}
Taking this into account we will neglect $A_2$ in what follows.

Now we rewrite the total transition amplitude  in the following,
more explicit form (retaining only the main contribution):
\begin{eqnarray}\label{zeeman-matrix-el-total+}
    \la f \vert \delta{\hat H} \vert i\ra^{(total)} &=& A \;
    \frac{v_{0}}{V}\;
    \sqrt{\frac{\hbar}{2\rho \, \omega_{\bf{k},\lambda} V_{ph}}}
    \;\sqrt{n_{\omega_{\bf{k},\lambda}}+1}
    \nn \\
    &\times& \sum_{j} \left(e^{\text{i}{\bf k}\cdot{\bf R}_{j}}
    - 1\right)
    \hat{S}^{\al}_{+-}\, \la f_N \vert \;\hat{
    I}^{\al}_j \vert i_N\ra \; \nn \\
    &\times& ({\bf \epsilon}_{\bf{k},\lambda}
    \cdot {\bf \nabla}_{{\bf R}_{j}}) \;
    \varphi_0^2 ({\bf R}_{j}),
\end{eqnarray}
where $\hat{S}^{\al}_{+-}=\la {\bf n}_{+}\vert \hat{S}^{\al}\vert
{\bf n}_{-}\ra$ (a sum over $\al$ in Eq.
(\ref{zeeman-matrix-el-total+}) is assumed).

By means of Fermi's golden rule we obtain the following expression
for the transition rate (including a sum over final states and a
proper average over initial states):
\begin{eqnarray}\label{zeeman-rate}
    \dot{W} &=& \frac{2\pi}{\hbar} \frac{A^2}{N^2}
    \int\frac{d^3{\bf k}}{(2\pi)^3}\;\delta (\hbar\omega_{{\bf k}}-\hbar\omega)
    \nn\\
    &\times& \frac{\hbar}{2 \rho \, \omega_{\bf k}}\,
    (n_{\omega_{\bf k}}+1)
    \sum_{j, j'} F({\bf R}_{j}, {\bf R}_{j'}) \; G_{jj'} \nn\\
    &\times& \left[\,{\bf \nabla}_{{\bf R}_{j}}
    \varphi_0^2 ({\bf R}_{j})
    \cdot {\bf \nabla}_{{\bf R}_{j'}}
    \varphi_0^2 ({\bf R}_{j'})\,\right],
\end{eqnarray}
where $N=V/v_0$ is the number of unit cells inside the dot volume
$V$. In (\ref{zeeman-rate}) we have employed the sound wave
dispersion law in the form $\omega_{\bf k} = k s$, i.e. we have
neglected  the difference in the transverse and longitudinal sound
velocities in summation over phonon polarizations. This simplifies
our formulas but should not change appreciably our numerical
results.

For clarity, in Eq. (\ref{zeeman-rate}) we have combined some
exponential factors from Eq. (\ref{zeeman-matrix-el-total+}) into
the following expression:
\begin{eqnarray}\label{exp-factor}
    F( {\bf R}_{j}, {\bf R}_{j'}) = 4 \,
    e^{\text{i}{\bf k}\cdot\frac{{\bf R}_{j}-{\bf R}_{j'}}{2}}
    \sin\frac{{\bf k}\!\cdot\!{\bf R}_{j}}{2}
    \sin\frac{{\bf k}\!\cdot\!{\bf R}_{j'}}{2}.
\end{eqnarray}
We have also separated all spin-related factors into the
correlation function (cf. Refs. \onlinecite{ENF, EN}):
\begin{eqnarray}\label{zeeman-spin-factor}
    G_{jj'}=\hat{S}^{\al}_{-+}\hat{S}^{\be}_{+-}
    \la i_N\vert\;\hat{I}^{\al}_j \,
    \hat{I}^{\be}_{j'}\vert i_N \ra _{av},
\end{eqnarray}
where the subscript, {\it av}, indicates averaging over initial
nuclear spin states  (in our case over a thermal distribution).

At temperatures much higher than $\sim 10^{-7}\, K$, which is the
order of magnitude of the nuclear spin-spin interaction, the
nuclear spins are not correlated, i.e. $G_{jj'}=G \,
\delta_{jj'}$. We suppose that there are no other sources of
average nuclear polarization either. This means, in turn, that the
interference terms in Eq. (\ref{zeeman-rate}), stemming from
different nuclei, $j\neq j'$, vanish.

With the help of the usual commutation rules for spin components
and the equality $\la {\bf n}_{-}\vert \hat{{\bf S}}\vert {\bf
n}_{-}\ra = -\frac{1}{2}{\bf n}$, we obtain for each nuclear spin
the formula
\begin{eqnarray}\label{spins-relation}
    \hat{S}^{\al}_{-+}\hat{S}^{\be}_{+-}\,\hat{I}^{\al} \,
    \hat{I}^{\be}=\frac{1}{4}\left[\hat{{\bf I}}^2-(\hat{{\bf I}}
    \cdot {\bf n})^2+ (\hat{{\bf I}}\cdot {\bf n})\right],
\end{eqnarray}
which results in a correlation function $G=\frac{1}{6}I(I+1)$,
provided the average nuclear spin is zero.

Taking all of this into account, the following expression
describes our main result for the rate of the electron-spin
relaxation between Zeeman sublevels of the ground state, due to
the hyperfine-phonon mechanism considered here:
\begin{eqnarray}\label{zeeman-t1}
    \frac{1}{T_{1}} &=& \frac{A^2}{N}
    \frac{n_{\omega}+1}{6 \pi\hbar \rho \, s^3}\;\omega \;I(I+1) \nn
    \\
    &\times& \overline{ \left[ 1 -
    \frac{\sin(k \vert{\bf R}'\vert)}
    {k \vert{\bf R}'\vert}\right]
    \left[{\bf \nabla}_{{\bf R}'}
    \varphi_0^2 ({\bf R}') \right]^2}.
\end{eqnarray}
The overbar in Eq. (\ref{zeeman-t1}) indicates an average over the
positions ${\bf R}'$ of all nuclei in the dot. If the electron
envelope wave function changes little on the scale of distance
between the nuclei, then this is just a spatial average over an
electron localization volume. In deriving Eq. (\ref{zeeman-t1}),
we have used the identity
\begin{eqnarray}\label{angle-average}
    \int \! d \Omega_{{\bf k}} \, \sin^2({\bf k}\cdot {\bf \xi})
    = 2\pi \left[1 - \frac{\sin(2 k \vert{\bf \xi}\vert)}
    {2 k \vert{\bf \xi}\vert}\right].
\end{eqnarray}

We note that the relaxation rate for an individual {\em nuclear}
spin is obtained by dividing Eq. (\ref{zeeman-t1}) by $N$.

\section{Triplet-Singlet transition}
\label{tripletsec}

We proceed in the same way in order to calculate the transition
rate for two electrons that initially reside in the lowest-lying
triplet state and decay towards the ground-state singlet. We
suppose that the Zeeman splitting (produced by external and/or
nuclear magnetic fields) can be neglected as compared to the
orbital energy spacing that defines the transition energy for this
process.

The wave function corresponding to the initial spin-triplet
electron state is
\begin{eqnarray}\label{st-initial}
     \vert i_e^{T\!r}\ra = \vert \psi^{T\!r}({\bf r}_1, {\bf r}_2)
     \ra\otimes \vert T\!r \ra,
\end{eqnarray}
where the coordinate wave function is assumed to be given by a
Slater determinant (i.e. neglecting correlations):
\begin{equation}\label{st-initial-coordinate}
    \psi^{T\!r}({\bf r}_1, {\bf r}_2) = \frac{\varphi_0({\bf r}_1)
    \varphi_1({\bf r}_2)-\varphi_0({\bf r}_2)\varphi_1
    ({\bf r}_1)}{V\sqrt{2}} \,.
\end{equation}
Here $\varphi_1({\bf r})$ is the wave function of the first
excited single electron orbital state normalized according to Eq.
(\ref{zeeman-initial-coordinate}). For concreteness, we choose it
to correspond to the quantum numbers $n=0$ and $\lambda=1$ (in an
external magnetic field the energy of this state is less than
$\varepsilon_{0, \,-1}$), and we note that there is no term in the
total Hamiltonian which couples directly states with $\lambda=\pm
1$. We write for the spin part (as in Ref. \onlinecite{ENF})
\begin{equation}\label{st-initial-spin}
    \vert {T\!r} \ra = - \frac{\nu_x - i\nu_y}{\sqrt{2}}\vert 1,+1\ra
    + \frac{\nu_x + i\nu_y}{\sqrt{2}}\vert 1,-1\ra + \;\nu_z \vert
    1,0\ra ,
\end{equation}
where the coefficients $\nu_{x,y,z}$ determine the initial
superposition of degenerate states $\vert S, m\ra$ with different
{\it z}-components $m=\pm 1, 0$ of the total spin $S=1$ of two
electrons.

For the final spin-singlet state we have
\begin{eqnarray}\label{st-final}
     \vert f_e^{Si}\ra = \vert \psi^{Si}({\bf r}_1, {\bf r}_2)
     \ra\otimes \vert Si \ra,
\end{eqnarray}
where $\vert Si \ra$ denotes the singlet spin state ($S=0$), and
the coordinate wave function is given by the expression:
\begin{eqnarray}\label{st-final-coordinate}
    \psi^{Si}({\bf r}_1, {\bf r}_2) = \frac{\varphi_0({\bf r}_1)
    \varphi_0({\bf r}_2)}{V} \,.
\end{eqnarray}

Again, the transition amplitude is given by Eqs. (\ref{zeeman-a1})
and (\ref{zeeman-a2}), where we have to introduce sums over
electron coordinates: e.g. $\nabla _{\bf r}\hat{V}_{h\!f}({\bf
r}\!-\!{\bf R}_{j}) \rightarrow \sum_i \nabla _{{\bf r}_i}
\hat{V}_{h\!f}({\bf r}_i\!-\!{\bf R}_{j})$, etc.

In the present case, the energy of the emitted phonon, $\hbar
\omega$, is equal to the single-particle energy splitting, $\hbar
\omega_0$. In the expression for $A_2$, Eq. (\ref{zeeman-a2}), the
contribution from the spin-singlet intermediate state dominates,
due to the small denominator [\onlinecite{ENF}] given by the
exchange splitting $\delta_{ST}$. However, although $\delta_{ST}$
is smaller than $\hbar \omega_0$, it still has the same order of
magnitude \cite{ENF}, and the ratio $\vert A_2/A_1 \vert \sim
(z_0/l)(\omega_0/\delta_{ST})$ is still much less than unity (here
we have $k l \gg 1$). Therefore we can once more neglect $A_2$.

We obtain for the triplet-singlet transition rate:
\begin{eqnarray}\label{st-rate}
    \dot{W}^{ST}\!\!\!&=&\!\! \frac{2\pi}{\hbar}
    \frac{A^2}{N^2} \int
    \frac{d^3{\bf k}}{(2\pi)^3}\;\delta (\hbar\omega_{{\bf k}}-\hbar\omega)
    \nn \\
    &\times& \!\!\frac{\hbar}{2 \rho \, \omega_{\bf{k}}}\,
    (n_{\omega_{{\bf k}}}+1)
    \sum_{j,j'}F({\bf R}_{j}, {\bf R}_{j'})\; G_{jj'}^{ST} \nn\\
    &\times& \!\! \frac{1}{2}\!\left[{\bf \nabla}_{{\bf R}_{j}}
    (\varphi_0 ({\bf R}_{j})\varphi_1 ({\bf R}_{j}))
    \cdot{\bf \nabla}_{{\bf R}_{j'}}(\varphi_0 ({\bf R}_{j'})
    \varphi_1 ({\bf R}_{j'}))\right]\!\!,\nn\\
\end{eqnarray}
where the correlation function is:
\begin{eqnarray}\label{st-spin-factor}
    G_{jj'}^{ST}&=& {\la T\vert\hat{S}^{\al}_1-\hat{S}^{\al}_2\vert S\ra
    \la S\vert\hat{S}^{\be}_1-\hat{S}^{\be}_2 \vert T\ra}
    \la i_N\vert\;\hat{I}^{\al}_j \cdot
    \hat{I}^{\be}_{j'}\vert i_N \ra_{av} \nn \\
    &=& \! {\nu^{* \al} \nu^{\be}}\la i_N\vert\;\hat{I}^{\al}_j \cdot
    \hat{I}^{\be}_{j}\vert i_N \ra_{av}\delta_{jj'}
    = \frac{1}{3}I(I+1)\delta_{jj'}.
\end{eqnarray}
Again, the nuclear spin state is averaged over a completely
unpolarized thermal distribution.

Finally, the relaxation rate in this case is:
\begin{eqnarray}\label{st-t1}
    \!\!\!\!\! \frac{1}{T^{ST}_{1}}\!\!\! &=& \! \frac{A^2}{N}
    \frac{n_{\omega}+1}{6 \pi\hbar \rho \, s^3}\;\omega \;I(I+1)
    \nn \\
    &\times& \!\!\! \overline{\left[ 1 -
    \frac{\sin(k \vert{\bf R}'\vert)}
    {k \vert{\bf R}'\vert}\right]\!\!
    \left[ {\bf \nabla}_{{\bf R}'}\left(\varphi_0 ({\bf R}')
    \varphi_1 ({\bf R}')\right) \right] ^2}.
\end{eqnarray}

\section{Numerical estimates and discussion}
\label{numestsec}

In order to estimate the rates in both of the cases that have been
considered above, we take into account realistic dimensions of
typical quantum dots. Usually the lateral length of the dot is
much larger than its transverse dimension, $l \gg z_0$, and we use
the following approximation: $\left[ {\bf \nabla}_{{\bf
R}'}\left(\varphi_0 ({\bf R}') \varphi_1 ({\bf R}')\right) \right]
^2 \simeq \left[ {\bf \nabla}_{{\bf R}'} \varphi_0^2 ({\bf R}')
\right]^2 \simeq \frac{1}{(z_0/2)^2}$ (which we suppose holds for
the average over nuclear positions ${\bf R}'$). For QDs with a
disc shape, in the typical limit $k z_0 < 1$, we can obtain a
simple analytical expression for the average,
\begin{equation}\label{sinc-average}
    f(k l)\equiv \overline{\left[ 1 - \frac{\sin(k \vert{\bf R}'\vert)}
    {k \vert{\bf R}'\vert}\right]}
    \simeq 1 - \frac{2(1-\cos (k l))}{(k l)^2},
\end{equation}
which can be well approximated by $\min \{(k l)^2/12, 1\}$. We
remark that this factor, which is present in the case of the
co-moving confining potential, is absent when the confining
potential position is fixed. This is the only difference in the
results for the relaxation rates in these two cases, and it
becomes important only in the limit of small phonon frequencies,
$k l \ll 1$.

We can write the relaxation rate between Zeeman sublevels and
triplet and singlet electron states using the same approximate
expression (the difference lies in the energy scale $\omega$):
\begin{equation}\label{t1-approximate}
    \frac{1}{T_{1}} \simeq \frac{2}{3}\, \frac{A^2}{N}\;
    \frac{\omega\, (n_{\omega}+1)}{\pi\hbar\,\rho \, s^3 z_{0}^2}
    I(I+1)\, f(k l).
\end{equation}

The linear dependence of the relaxation rate on the phonon
frequency (for phonon wave lengths smaller than the size of the
QD, i.e. for sufficiently strong magnetic fields in the case of
Zeeman sublevels relaxation) sets our mechanism apart from those
considered earlier \cite{PBS, EN}, where a cubic phonon frequency
dependence is expected for low temperatures, $k_B T \ll \hbar
\omega$ (and a quadratic one in the opposite limit). In addition,
our result does not depend either on the proximity of the nearest
level, in contrast to the spin relaxation rates calculated in
Refs. \onlinecite{ENF, EN}.

The result for the electron-nuclei flip-flop transition rate
obtained in Ref. \onlinecite{PBS} is larger than our result, Eq.
(\ref{t1-approximate}), by a factor $\sim (\gamma z_0 / l)^2$
(provided $k l \ll 1$), which is of the order of $10$ ($\gamma
\simeq 50$) for typical QD sizes cited below, but can be smaller
for large QDs. We should be cautious, however, in applying
directly the reasoning of Ref. \onlinecite{PBS} to the
triplet-singlet transition, when the emitted phonon energy
corresponds to the electron binding energy and hence the condition
of adiabatic electron motion in a vibrating lattice (used in that
work) will not be fulfilled.

In GaAs all nuclei have spin $I=\frac{3}{2}$, and $A$ is the sum
of $A_j$ over all the nuclei in the unit cell: $A=\sum_j A_j
\simeq 90$ $\mu$eV \cite{DPreview}. The mass density is $\rho
\simeq 5.32 \times 10^3$ kg/m$^3$, and we approximate the mean
sound velocity by the velocity of transverse sound waves, $s \sim
s_t \simeq 3 \times 10^3$ m/s \cite{Adachi}. The typical
transverse dimension of a quantum dot is $z_0 \simeq 10$ nm and
its lateral size is $l \simeq 100$ nm. The dot contains about $N
\sim 10^5$ unit cells (8 nuclei in each). Hence, we can write for
the relaxation rate the approximate expression:
\begin{eqnarray}\label{t1-numerical}
    \frac{1}{T_{1}} \simeq 1 \times 10^{-15}\frac{\omega}
    {1 - \exp(-\frac{\hbar\,\omega}{k_B T})}
    \, f(k l).
\end{eqnarray}

\begin{figure}[t] \centering
\includegraphics[viewport=10 20 298 240, width=0.50\textwidth]{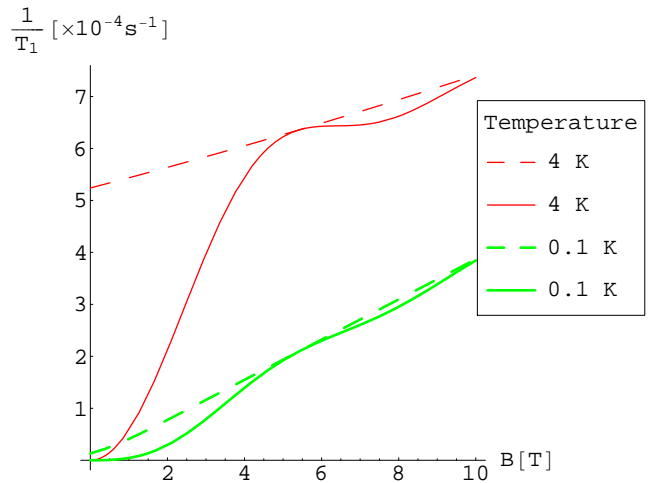}
\caption{The single electron spin relaxation rate in a GaAs QD
versus magnetic field (according to Eq. (\ref{t1-numerical})), derived
for the hyperfine-phonon mechanism discussed in the present article. The
solid line corresponds to the case when the confining potential
vibrates together with the lattice, while the dashed line refers
to the case of a fixed confining potential.}
\label{fig:frequency-dependence}
\end{figure}

For the transition between Zeeman energy sublevels the phonon
energy is equal to the Zeeman splitting, $ \vert \text{g}^* \mu _B
B\vert$, which corresponds to $0.025$ meV T$^{-1}$ in GaAs where
$\text{g}^* = - 0.44$. At the same time the cyclotron energy,
$\hbar \omega_c$, is as much as $1.76$ meV T$^{-1}$ due to the
small effective mass in GaAs, $m^*=0.067 m_e$. Consequently, the
condition $\omega \ll \omega_0$ is always satisfied when a
perpendicular magnetic field is applied. For in-plane magnetic
fields $\omega \ll \omega_0$ when $B \ll 10$ T in a lateral and $B
\ll 100$ T in a vertical QD (the single particle energy spacing in
a lateral QD is 100 - 300 $\mu$eV, and in a vertical QD the
confining energy of an approximate 2D harmonic potential is $\sim
4$ meV \cite{Kouwenhoven, FujiPhys}). The condition $k l \sim 1$
corresponds to a magnetic field of the order of $1$ T. Thus, for
magnetic fields larger than $\sim 1$ T the relaxation rate is
linear in the emitted phonon frequency and as a consequence it is
linear in the strength of the magnetic field (see Fig.
(\ref{fig:frequency-dependence}) which is plotted for the limit $k
z_0 < 1$ corresponding to $B < 10$ T). For the fixed or
"independent" confining potential we would have $f(kl)\equiv 1$
and the dependence of the rate on the emitted phonon frequency
would be linear even for small magnetic fields, i.e. when $k l \ll
1$ (apart from the temperature-dependent factor $(n_{\omega}+1)$).

As to the order of magnitude, the mechanism that has been
considered in this paper gives a rate of the order of $10^{-4}$
s$^{-1}$ for a temperature of about  1 K and a magnetic field of
about 1 T. It therefore appears to be much less efficient than the
piezoelectric coupling mechanism considered in Ref.
\onlinecite{EN}, where the rate is about 1 s$^{-1}$ for comparable
values of temperature, 4 K, and magnetic field, 0.5 T.

For the triplet-singlet transition in a lateral dot we find a rate
of the order of $10^{-4}$ s$^{-1}$ for temperatures up to $\sim 1
\, K$ ($\hbar \omega \sim 100 \mu$eV), and in a vertical QD we
have $1/T_1 \sim 10^{-3}$ s$^{-1}$ for temperatures up to $\sim
10$ K and $\hbar \omega \sim 1$ meV (we note that in both cases $k
l \gg 1$). The latter result should be compared with $1/T_1
\approx 2\times 10^{-2}$ s$^{-1}$ calculated in Ref.
\onlinecite{ENF}.

Let us now turn to spin relaxation in silicon. Taking into account
the natural abundance of $^{29}$Si nuclei with non-zero spin
$n_{I=1/2}= 4.68 \%$, their magnetic moment $\mu_{I} = -0.56 \,
\mu_N$, the lattice constant $a = 5.43$ \AA \hspace{1pt} and the
electronic density at the position of the nucleus $\eta \simeq
186$ \cite{Abragam} we find the effective hyperfine coupling
constant to be $A \simeq 5$ $\mu$eV. This is far smaller than in
GaAs, due to the smaller $\eta$ and smaller percentage of nuclei
with spin. Inserting the mass density of Si, $\rho \simeq 2.3
\times10^3$ kg/m$^3$ and the transverse sound velocity $s_t \simeq
5.4 \times 10^3 m/s$, we obtain a prefactor of the order of
$10^{-20}$ in Eq. (\ref{t1-numerical}). This corresponds to a very
small relaxation rate $1/T_1 \sim 10^{-9}$ s$^{-1}$ between Zeeman
sublevels in a magnetic field of $B =1$ T, for temperatures up to
1 K (g = 2 in Si). Note that the spin-orbit related electron spin
relaxation time in lateral Si QDs \cite{Glavin}, for Si:P bound
electrons and for QDs in SiGe heterostructures \cite{Tahan} has
been predicted recently to be on the order of several minutes for
the same values of magnetic field and temperature. In the latter
case, however, the relaxation rate may be strongly decreased,
below the values found for our mechanism, by application of
uniaxial compressive strain \cite{Tahan}.

In general, electron spin relaxation induced by electron-nuclei
hyperfine interaction in a quantum dot is not as efficient as
relaxation due to spin-orbit interaction for typical values of
system parameters \cite{KhN1, KhN2, WRL-G, MKGB, Tahan, Glavin,
ENF}. In many experiments, the hyperfine related rate would be
masked by the spin-orbit mechanism (However, see Ref.
\onlinecite{EN} for a case where the hyperfine mechanism may
dominate). The present experimental data for the triplet-singlet
transition rate in a GaAs vertical QD is $T_1 \approx 200 \mu$s at
temperatures up to 0.5 K and triplet-singlet energy splitting
$\varepsilon_{S-T} \sim 0.6$ meV \cite{FujiNature}.  For the
relaxation time between Zeeman sublevels in a lateral GaAs QD only
a lower bound is available: $T_1 \gtrsim$ 50 $\mu$s for an
in-plane magnetic field B=7.5 T at T=20 mK \cite{Hanson}. Both of
these measurement results were obtained by means of transient
current spectroscopy \cite{FujiPRB}.

On the other hand, hyperfine coupling mechanisms (such as the one
considered in this paper) may be particularly relevant as far as
effects like dynamic nuclear polarization are concerned, where the
electron-nuclei hyperfine interaction plays a crucial role
\cite{Abragam, DPreview, PBS}. Recently, a hyperfine nuclear spin
relaxation time on the order of 10 min was measured in a single
GaAs QD, at a bath temperature of 40 mK and a magnetic field up to
0.5 T \cite{Huttel}. However, this experiment dealt with a
nonequilibrium transport situation, with a resulting spin-flip
mechanism whose rate turns out to be orders of magnitude larger
than the one discussed in the present article (see Ref.
\onlinecite{Lyanda-Geller}).

\section{Conclusions}
\label{conclusions}

In summary, we have considered a specific mechanism for inelastic
electron spin relaxation in a QD induced by the electron-nuclei
hyperfine interaction in combination with lattice vibrations. We
have found that the interference between first and second orders
of perturbation theory is essential for a correct description of
the suppression of relaxation at small transition frequencies. The
relaxation rate has been calculated both for the transition
between Zeeman sublevels of the orbital ground state and for the
triplet-singlet transition. We have obtained estimates based on
these general results and realistic system parameters. The
estimates demonstrate that the relaxation rate due to this
particular mechanism is very small: For the spin relaxation
between Zeeman sublevels it is much less than the rate calculated
earlier in second order perturbation theory with an emission of a
phonon through piezoelectric electron-phonon coupling in a GaAs QD
\cite{EN}, and it is less than (or at most comparable with) the
relaxation rate due to the change in the localized electron
effective mass induced by the lattice dilation in silicon
\cite{PBS}. For the triplet-singlet transition, the rate in GaAs
QDs is still smaller by an order of magnitude than the
corresponding rate of the piezoelectric mechanism.

\smallskip

We would like to thank B. L. Altshuler, D. V. Averin, C. Bruder,
E. V. Sukhorukov and A. V. Chaplik for useful discussions.


\end{document}